# High-resolution dielectric characterization of minerals: a step towards understanding the basic interactions between microwaves and rocks

T. Monti[1*], A. Tselev[2], O. Udoudo[1], I. N. Ivanov[2], S. W. Kingman[1]

[1] Faculty of Engineering, University of Nottingham, NG7 2RD, Nottingham, UK

[2] Center for Nanophase Materials Sciences, Oak Ridge National Laboratories, Oak Ridge, Tennessee, 37831, USA

Abstract

Microwave energy has been demonstrated to be beneficial for reducing the energetic cost of several steps of the mining process. Significant literature has been developed about this topic but few studies are focused on understanding the interaction between microwaves and minerals at a fundamental level in order to elucidate the underlying physical processes that control the observed phenomena. This is ascribed to the complexity of such phenomena, related to chemical and physical transformations, where electrical, thermal and mechanical forces play concurrent roles. In this work a new characterization method for the dielectric properties of mineral samples at microwave frequencies is presented. The method is based upon the scanning microwave microscopy technique that enables measurement of the dielectric constant, loss factor and conductivity with extremely high spatial resolution and accuracy. As opposed to conventional 'bulk' dielectric techniques, the scanning microwave microscope can then access and measure the dielectric properties of micrometric-sized mineral inclusions within a complex structure of natural rock. In this work a 5 by 20 μm size hematite inclusion has been characterized at a microwave frequency of 3 GHz. scanning electron microscopy/energy-dispersive x-ray spectroscopy and confocal micro Raman spectroscopy were used to determine the structural details and chemical and elemental composition of mineral sample on similar scale.

1. Introduction

Microwave heating has been demonstrated to be potentially beneficial in reducing the energetic impact of various mineral processing unit operations (Haque, 1999), thereby making the mining process more efficient and sustainable (Kingman, 2006). For example, the effectiveness of the microwave-induced thermally assisted liberation has been extensively demonstrated (Walkiewicz et al., 1989). Microwave fields induce volumetric and selective heating, in a heterogeneous material the lossy phases only absorb electromagnetic energy, avoiding the waste through the 'bulk-heating' of the surrounding rock (Jones et al., 2005). Due to the different expansion rates of the minerals in the rock, micro-fractures are induced, theoretically making the grinding process less energy-consuming (Kingman and Rowson, 1998).

In spite of its potential, the application of microwave energy in the mining process has not delivered the outcomes required in order to become a commonly used technology in this field. The reasons are multiple but certainly can be ascribed to a lack of fundamental understanding of the interacting mechanisms between minerals and the microwave field itself. In particular, few pioneering studies have been proposed for correlating physical effects (mechanical, thermal) to the electromagnetic energy absorbed (Jones et al., 2005; Ali and Bradshaw, 2009; Whittles et al., 2006), and a sparse literature base has been developed so far for correlating these effects to the electromagnetic properties that are the most important for understanding the microwave-matter interactions.

The heating mechanism of minerals in electromagnetic fields is correlated to complex physical phenomena, related to dielectric and magnetic losses:

$$P_{av} = \omega \varepsilon_0 \varepsilon''_{eff} E^2_{rms} V + \omega \mu_0 \mu''_{eff} H^2_{rms} V \quad (1)$$

where $P_{av}$ is the average power deposited within the sample in watts [W], $\omega = 2\mu f$ is the wave angular frequency in hertz [Hz], $\varepsilon_0$ and $\mu_0$ are the vacuum permittivity and permeability, respectively in farads per metre [F/m] and henry per metre [H/m], $\varepsilon''_{eff}$ and $\mu''_{eff}$ are the electric and magnetic effective loss factors, $E_{rms}$ and $H_{rms}$ are the average intensities of the electric and magnetic field, respectively in volts per metre [V/m] and amperes per metre [A/m], $V$ is the volume of the heated sample in [m$^3$].
The deposited power is therefore dependent upon the material characteristics which are variable with frequency (Metaxas and Meredith, 1983):

$$\varepsilon''_{eff} = \varepsilon'' + \sigma/\omega\varepsilon_0 \quad (2)$$
$$\mu''_{eff} = \mu''$$

It is then essential to evaluate the complex dielectric permittivity $\varepsilon^* = \varepsilon' - j\varepsilon''_{eff}$ and magnetic permeability $\mu^* = \mu' - j\mu''_{eff}$ of the sample under analysis. The first quantity, at microwave frequencies, is mainly determined by the polarization effect, quantified by $\varepsilon''_{eff}$. The only exception is for highly conducting materials, where the conduction phenomenon, quantified by σ, is dominant. $\mu''_{eff}$ is usually negligible for non-magnetic samples.

The heterogeneity of ores makes the interaction with the microwave field highly complex. Selective heating of certain mineral inclusions with respect to the surrounding matrix typically occurs (Kingman et al., 2000; Walkiewicz et al., 1988). This is because such mineral inclusions have significantly higher dielectric properties than the rest of the rock forming material, namely a higher $\varepsilon''$. From Eq. (1), the power absorbed by the mineral inclusions is therefore higher than the surrounding gangue material, so the microwave energy 'focuses' selectively on such inclusions, giving rise to a temperature differential which is the basis of many applications of microwave energy in minerals processing.

The dielectric properties of each phase are required to describe such a complicated selective heating phenomenon from an electromagnetic point of view and then relate it to a multi-physics description of the process. Unfortunately, conventional methods for the dielectric characterization of materials are not suited to this as they allow only the quantification of the bulk properties of the material as opposed to the individual phase which constitute the bulk.

Different techniques have been applied for mineral characterization in the Ultra High Frequency (UHF) microwave range (300 MHz – 3 GHz), where the Industrial-Scientific-Medical (ISM) spectra are located. For example, the short circuited waveguide method, based on the previously proposed standing wave perturbation method by Roberts and Von Hippel (1946), is used for the analysis of minerals between 1 and 22 GHz (Nelson et al., 1989). In Tinga (1989), a bridge configuration, with samples mounted in the central section of a waveguide, has been used for testing the dielectric properties of some metal oxides at 2.45 GHz over a wide range of temperature. An open-ended coaxial line technique has been widely exploited as well: in reference Salsman and Holderfield (1994) chalcopyrite, chalcocite and cobaltite were analysed on a limited temperature range [0-300 C].

The main limitation of these 'bulk' techniques is that they cannot quantify the properties of the individual minerals as they are in the natural mixture that forms the rocks. Additionally, the majority of conventional dielectric measurement techniques are highly dependent on the density of the sample under study (Nelson et al., 1989). Application of mixture theories is then needed for calculating the dielectric properties of bulk samples.

In this way the only possible approach for simulating and testing a realistic structure was the one adopted in Salsman et al. (1996) which involved bulk measurements of both the thermal and dielectric properties of the constituent minerals. Bulk minerals were then crushed and sized to a certain mean particle size and then mixed in varying proportion creating a 'controllable' synthetic compound.

In this paper a novel method for measuring the dielectric properties of mineral inclusions embedded in a natural rock at microwave frequencies is presented. Such a method is based upon Scanning Microwave Microscopy (SMM) that is able to achieve extremely high spatial resolution and accurately measure the dielectric constant and losses of the sample under analysis.

Detailed numerical models, as well as more realistic and comprehensive multi-physics simulations of the actual systems, can then be carried out. Furthermore, the SMM can be used for quantifying the dielectric properties of treated samples and relate them to changes in physical structure, phases and chemical composition at the nanometric scale.

Similar studies, where dielectric properties are related to materials composition, were performed on homogeneous samples by measuring the dielectric properties with conventional 'bulk' techniques (Yoon et al., 2006).

In this work the SMM technique is introduced and exemplified for the first time to a hematite inclusion embedded in a gangue matrix. A comprehensive chemical and dielectric characterization of such a micrometric inclusion is shown. The SMM capability of mapping the dielectric properties with sub-micrometric resolution is highlighted.

2. Materials and methods

A set of ore samples were core drilled to produce sample mounts of 25mm diameter and 15mm height. The mounts were cleaned with distilled water and dried using compressed air. The cleaned mounts were polished using a Struers Rotopol polishing machine to a surface roughness of 10nm, necessary for achieving the finish required by both the scanning microwave and Raman microscopes.

In Figure 1 adjacent images of a portion of one of the sample surfaces is shown. For example, 5X magnification is used for the left 'montage' map (6 x 5 mm area); a 20X magnification is used for the right one (1.5 x 1.2 mm area), so that it is possible to have an understanding of the complex topographical scenario and related complex chemical and elemental composition of the samples under analysis.

The different grey tones are ascribable to the different materials that compose the rock. It is possible to identify a large number of phases in both of the images, with random shapes and positions. This means that the phases have different sizes and that they are effectively mixed to give a complex heterogeneous compound.

As it is necessary to chemically characterize this compound at a single-phase level in order to have a comprehensive description of the composition, it is therefore, required to have an instrument that characterizes the dielectric properties at similar level.

No further elements extraneous to the selected rocks were added during sample preparation in order to exclude any possible uncertainty due to contamination.

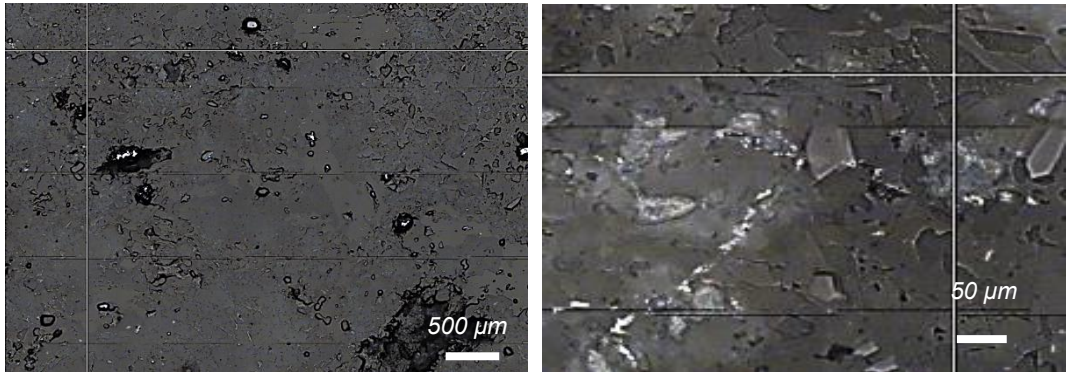

Fig.1 Optical images of a natural rock sample surface. The image is composed of several sub images stitched to form final image. A high heterogeneity of the surface is visible at different scales. Different grey tones are ascribable to different materials. Phases are clearly recognizable at very different scales.

2.1. Micro-Raman Spectroscopy

In this work, the micro-Raman spectroscopy technique (Colthup et al., 1975) is used for obtaining an elemental characterization of the sample at sub-micrometric scale. Since the spatial resolution is determined by the width of the spot of the probing light, it is possible to get a resolution that is close to that of the SMM.

Raman spectra are collected in backscattering geometry using a Renishaw 1000 Raman confocal microscope through a 50 objective (Leica, NA = 0.75) using a 532 nm laser as the excitation source. The Raman spectra were collected in the 100 to 1000 $cm^{-1}$ frequency range by scanning the area of interest in 2-0.6 micron steps point by point. Raman spectra of pure minerals were acquired on the same Raman spectrometer to assign inclusions to a particular mineral and to enable deconvolution of Raman spectra (Collins et al., 2014).

2.2. SEM-EDX Elemental Analysis

Micro-Raman spectroscopy and SEM-EDX techniques are complementary and allow accurate elemental and chemical characterization of the sample constituents since they are based on different probing mechanisms. They can both determine the chemical composition with micrometric and sub-micrometric resolution that can be directly compared to the SMM dielectric maps.

Scanning Electron Microscopy (SEM) – Energy Dispersive X-ray (EDX) spectroscopy characterisation was undertaken using the FEI Quanta 600i SEM equipped with 2 Bruker XFlash 5030 detectors. The EDX information was obtained using the Bruker Espirit version 1.9 software designed by Bruker Nano GMBH. The software is equipped to produce elemental information in the form of mineral maps (maps of the various elements identified in the sample via x-ray microanalysis) and elemental spectra from which the mineralogical composition is determined. The mineral maps are presented later.

The SEM provides backscatter (BSE) images of samples from which different mineral phase can be delineated. This information is presented in the form of grey scale images with brightness intensity values ranging from 0 to 255, where 0 represents black and 255 white. For the ore the BSE brightness for the copper sulphide and iron oxide minerals are over 150 and appear as bright features against the much darker gangue (see Figure 2). With these feature differences the phase of interest is segmented and x-rays collected from this phase where used to characterise them.

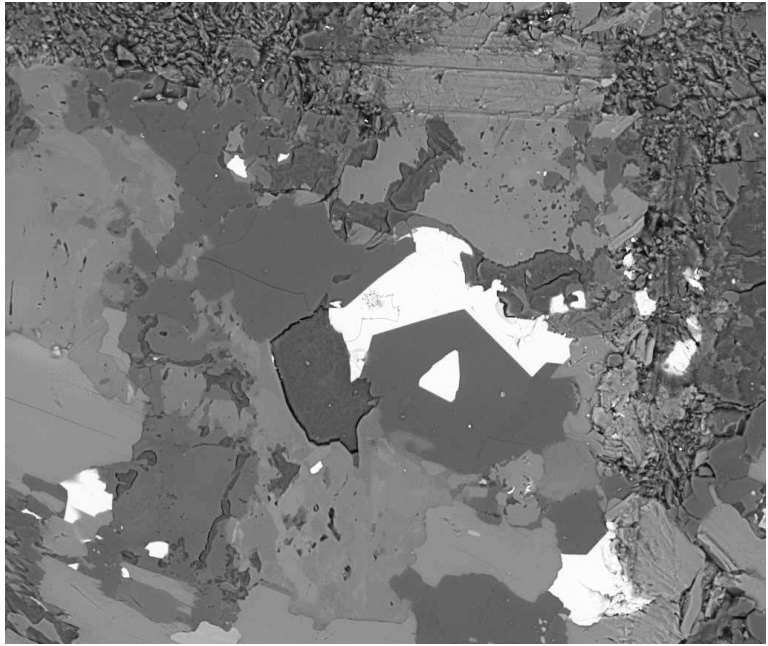

*Fig. 2 A BSE image showing brightness intensities of mineral of interest and gangue. The minerals appear significantly brighter than the darker gangue.*

*2.3. Scanning Microwave Microscopy (SMM)*

The SMM is one among the family of Scanning Probe Microscopy (SPM) techniques, where a sharp probe records a change in a certain physical quantity due to interaction with a sample. This class of techniques has been intensively developed since the invention of a Scanning Tunnelling Microscope (STM) in 1982 (Binnig and Rohrer, 1983) and became exponentially important because of its suitability for nanometre scale analysis. One of the best known SPM techniques is the Atomic Force Microscopy (AFM) (Binnig et al., 1986). The main feature of SPMs is an extremely high resolution that can be achieved while quantitatively measuring certain physical properties. SPM techniques are able to record short-range physical properties and their changes for each point of the scan. Through proper transducers and electronics, it is possible to obtain images of the scanned samples related to the physical interactions of interest.

The SMM records the variation of the microwave field emitted and reflected by a sharp tip placed either in close proximity or in contact with the sample. Due to the rapid decay of the interacting field and the extreme proximity between tip and sample, the range of interaction is within the so called 'reactive near-field' region of the emission. It is then possible to overcome the conventional resolution limits associated with far-field imaging and to obtain a sub-wavelength resolution (Anlage et al., 2007).

In this work, a commercial SMM system from Prime Nano Inc.© (http://www.primenanoinc.com/) installed on an Asylum Research MFP-3D AFM platform was used.

As described in the scheme in Figure 3, a microwave reflectometer, tuned at a working frequency $f = 3 GHz$, records in two separate channels the variations in the reactive (sMIM-C) and resistive (sMIM-R) parts of the tip-sample admittance (fig. 4). The signal change in the sMIM-R channel is determined by the sample conductivity and increases when such conductivity (loss) increases. In order to separate the two measurement channels and correspondingly, the two physical contributions, an initial calibration is needed on a calibration grating made of metallic patches on an $SiO_2$/Si substrate. Topographical AFM images of the samples are obtained simultaneously with the SMM characterization.

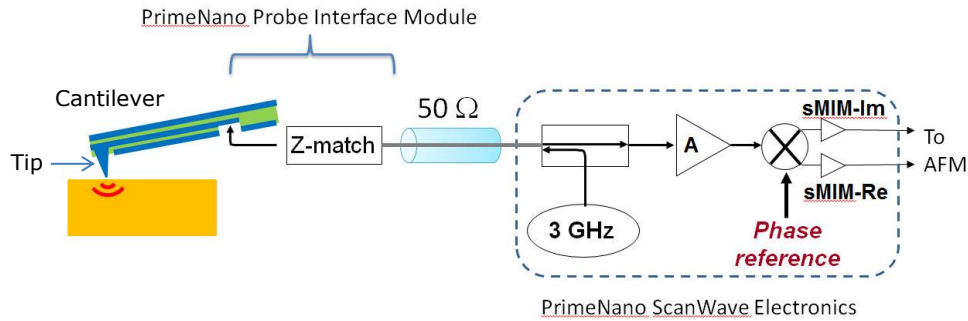

*Fig. 3 Schematic of the SMM based on Prime Nano Inc.©. The yellow sample is scanned by the probe that is connected to a single frequency microwave source. The microwave signal reflected back from the sample is carried on the tip cantilever from the apex of the tip to the 'z-match' circuit. The electronics is then able to separate the sMIM-Im and sMIM-Re signals.*

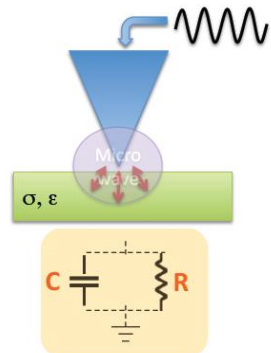

*Fig. 4 Lumped elements equivalent circuit description of the SMM tip-sample interaction mechanism: the reactive interaction (modelled by a parallel capacitance) depends on the ε' of the sample, while the resistive interaction (modelled by a parallel resistance) depends on σ.*

In order to obtain quantitative data from the scanned samples, another calibration procedure is needed, in particular, with this latter calibration it is possible to take into account the change in the tip apex shape during the measurements. It is particularly important in this work since the tip is in contact with the sample surface while scanning, and therefore, it is subject to a small, but continuous, alteration. Therefore, the latter calibration has been performed before and after each set of scans, while the former calibration on the grating was needed only when a probe was replaced with another.

Five different standard calibration samples with known dielectric permittivity were chosen for the latter calibration procedure: quartz $SiO_2$ (ε'≈4), magnesium oxide MgO (ε'≈10), lanthanum aluminate $LaAlO_3$ (ε'≈24), titanium dioxide $TiO_2$ (ε'≈100), strontium titanate $SrTiO_3$ (ε'≈300). Each standard sample was a few hundred microns thick. The absolute values of the sMIM signals are subject to small drifts, which can, however, significantly affect the measurement results. Therefore, all measurements are performed in 'differential' mode with use of the so-called 'NAP mode' of the AFM microscope (Asylum Research MFP-3D). In this way, numerical values from the sMIM channels are always referenced to the "in air" condition with the tip lifted 5 μm away from the sample surface. During measurements, each row of the scan is repeated twice: once with the tip in contact with the sample and once with the tip raised above the sample. The values of ΔR and ΔC below referrer to these differences between contact and "in air" conditions for both microscope channels.

By measuring the sMIM-C values related to the "in air" reference for the standards and then fitting them, it is possible to obtain an empirical calibration curve for the C-channel. Being mainly related to the reactive interaction of the microwave field with a sample, these values are influenced only by the real part of the sample dielectric permittivity (ε'). An example of such a calibration curve is shown in Figure 5.

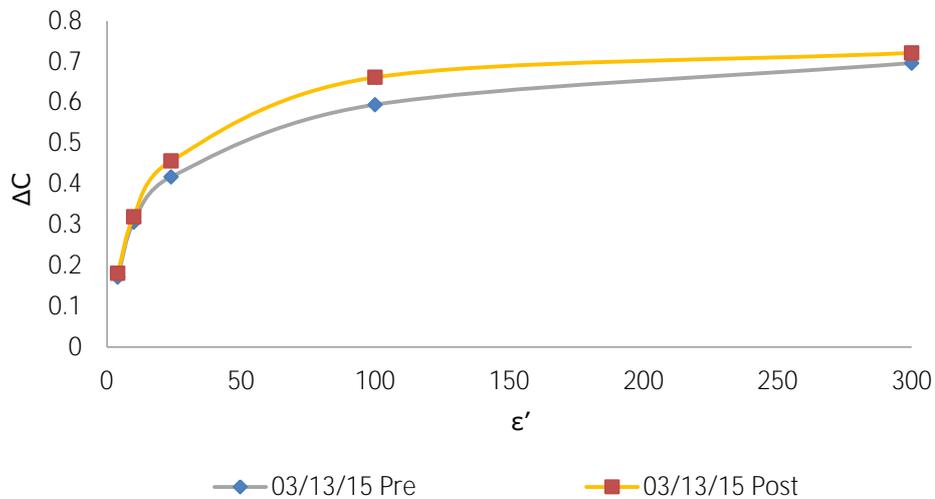

*Fig. 5 Example of a calibration curve for ε' data, interpolating the values recorded by the C-channel of the microscope referred to the 'in air' condition (ΔC) on dielectric known samples (SiO$_2$ (ε'≈4), MgO (ε'≈10), LaAlO$_3$ (ε'≈24), TiO$_2$ (ε'≈100), SrTiO$_3$ (ε'≈300). The grey curve was measured before the rock analysis, the yellow one was obtained afterwards. The discrepancy between the two curves reflects the changes in size and shape of the SMM probe due to the contact with the sample during the scanning process.*

3. Semi-empirical model of the SMM interaction

In order to obtain quantitative measurements for the resistive R-channel which is influenced by the loss tangent *tanδ* of a sample, it is not straightforward to create a calibration curve through measuring 'known loads'. Natural materials are obviously not well-characterized in terms of loss tangent since this property is influenced by a number of factors. Additionally, it is impossible to obtain well-defined calibration standards with proper combinations of permittivity and loss tangent, especially around the values of the naturally occurring mineral phases within an ore. In this work, we developed a semi-empirical calibration method for estimating the loss tangent of samples. Namely, we model the tip-sample interaction using a *quasi-static* numerical model in an AC/DC module of a finite-element analysis (FEA) package COMSOL Multiphysics©. The reference AC voltage of amplitude $V_{ac} = 1$ and a frequency $f = 3 GHz$ is applied to the tip in the model against a ground electrode simulating the sMIM probe shield. Figure 6 shows the geometry of the model, where ground boundary conditions are applied to the AFM cantilever as indicated by the red arrow.

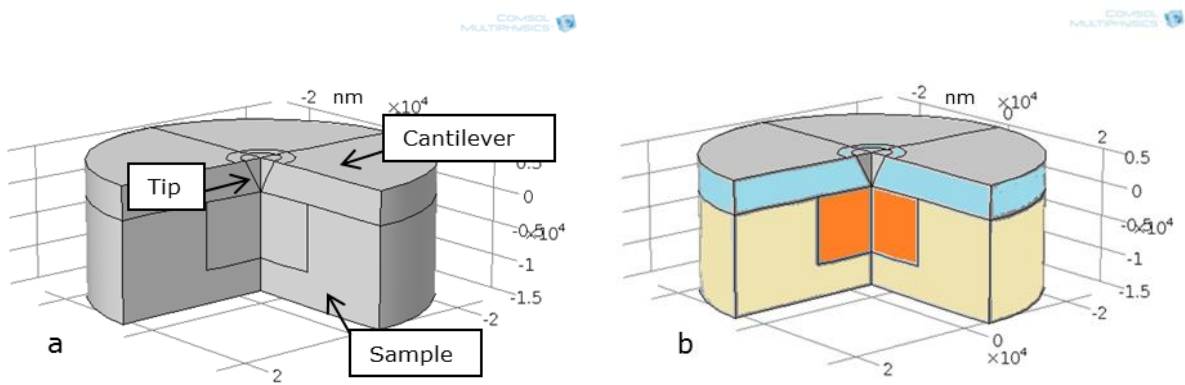

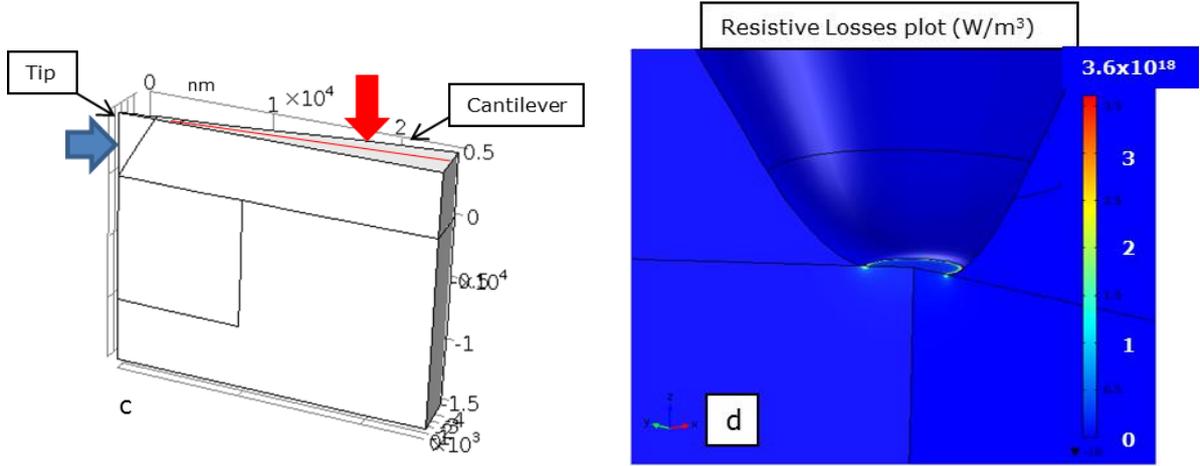

*Fig. 6a. Simplified geometry of the system implemented in COMSOL Multiphysics©. The actual finite-elements model is 2D axisymmetric. The 3D representation is obtained by rotating the section in Fig.6c. 6b.The grey represents the metallic parts of the tip and cantilever. The cyan represents air gaps between the sample surface and the cantilever. The orange represents an inclusion of a certain material different from the surrounding (cream). 6c.The red arrow and line indicated where the ground condition is applied, while the reference voltage $V_0$ is applied to the tip (blue arrow). 6d. A magnified view of the geometrical contact between tip-sample that well represents the real physical interaction. The simplified shape of the cantilever, circular instead of rectangular, is due to the rotation operation. Anyway this does not significantly affect the results of the model in respect to the tip-sample interaction.*

A quasi-static approximation is applicable since the dimensions of the model—a few micrometers—are small compared to the wavelength of the electromagnetic radiation in vacuum at 3 GHz, which is 10 cm. In the simulation, the permittivity of the sample is set to the same as the measured one and the conductivity of the sample is varied. The model is solved for the complex impedance *Y* of the tip-sample system as a function of the sample conductivity *σ*. The ratio of the real and imaginary parts of the calculated admittance Im(*Y*)/Re(*Y*) corresponding to the measured ratio of the sMIM signals $\Delta R/\Delta C$, where $\Delta R$ is the signal recorded by the R-channel of the microscope and the $\Delta C$ is recorded by the C-channel as described above and yields the sample conductivity.

The obtained value of *σ* is further used to obtain the sample loss tangent with the expression:

$$tan\delta = \frac{\sigma}{2\pi f \varepsilon_0 \varepsilon\prime} = \frac{\varepsilon\prime\prime}{\varepsilon\prime} \, , (3)$$

where $\varepsilon_0$ is vacuum permittivity.

The admittance is a function of a number of factors that are directly derived from the sMIM measurements such as the size of the objects of interest and their dielectric constant. As a starting value of the expected conductivity of the samples, loss tangent and conductivity values found in the literature (Nelson et al., 1989; Peng et al., 2012; Hotta et al., 2010) were used, after the samples were characterized by Raman and SEM-EDX techniques in terms of chemical and elemental composition. Since the tip-sample system can be considered as a lossy capacitor with the main contributions to the capacitance and loss variations stemming from the sample portion right beneath the tip-sample contact for semi-empirical estimations we can apply the relationship:

$$tan\delta = \kappa \frac{\Delta R}{\Delta C} \, , (4)$$

where the factor *κ* is insignificantly dependent on sample permittivity for large enough samples of about three times the tip pyramid height of ca. 5 μm. This fact can be used to generate loss tangent maps based on permittivity maps using a small set of the simulations, which effectively "calibrate" the factor *κ*.

4. Results and Discussion

*4.1. SEM-EDX Analysis and Confocal Micro-Raman Spectroscopy*

The set of techniques described so far have been applied for a comprehensive nanoscale characterization of natural rock constituents. In particular, the combined analysis through micro-Raman and SEM-EDX techniques, allowed for recognizing the chemical composition of the samples under analysis. The nature of the mineral inclusions has been determined with nanometric resolution, as detailed in the following pictures. Two hematite inclusions within a complex dielectric matrix have been tested as exemplar samples for the SMM technique.

From the SEM-EDX analysis (see Fig.7) we could define the exact elemental composition of the mineral inclusions and the surrounding matrix. The images clearly highlight the presence of two spots whose composition is essentially Iron Oxide (Fe+O). Their size is less than 30um.

They are embedded in a complex structure mainly identified as Calcite (Ca+C) with some Mg and Al contamination. Everything is surrounded by a wider quartz matrix (Si+O).

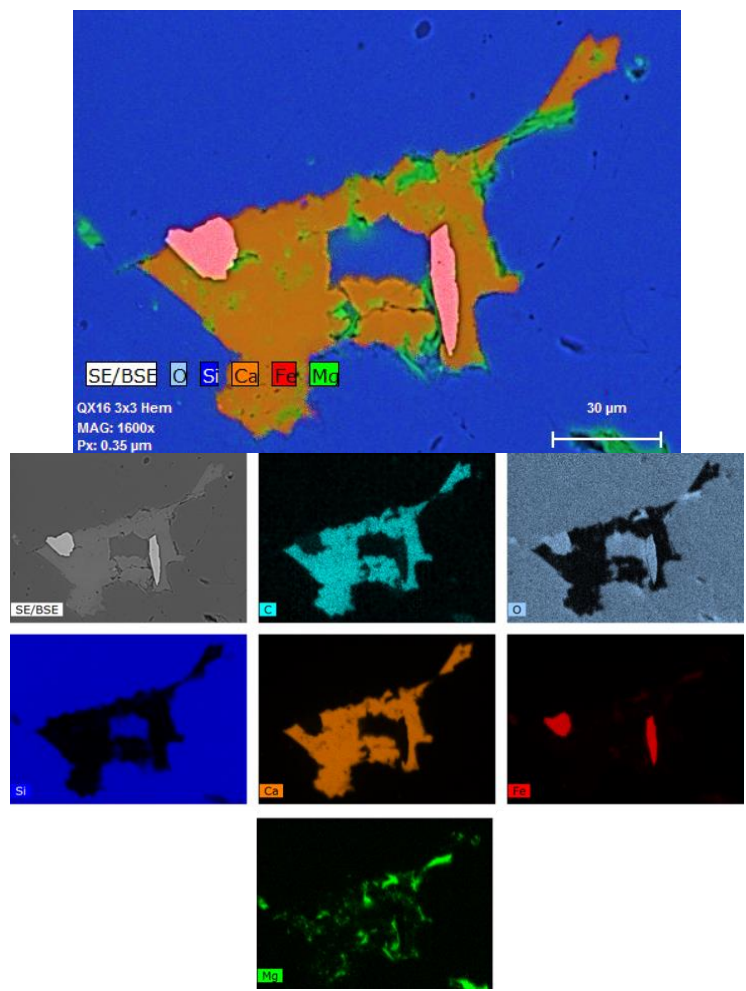

*Fig.7 SEM/EDX analysis of the hematite inclusions and their surrounding dielectric gangue matrix. The top image collects the contributions coming from the different chemical elements by overlapping the different colours used for them. The 'pink' hematite is then the combination of the 'red' iron and the 'white/grey' oxygen. The two inclusions are just a few tens of micrometres and they are chemically uniform. They are surrounded by a larger phase composed of 'brown' calcium carbonate (calcite), a combination of 'orange' calcium and 'cyan' carbon. This contains Mg and Al impurities. Everything is surrounded by 'light blue' quartz ('blue' silicon + 'white/grey' oxygen).*

A smaller area of one of the two hematite inclusions was investigated using confocal micro Raman in order to univocally determine the kind of iron bearing oxide, since hematite, magnetite and goethite have distinct Raman spectra where it is more complicated to distinguish them with the SEM/EDX analysis only. The confocal micro Raman technique can be used to record the Raman spectra of the materials with a sub-micron diffraction limited resolution (Collins et al., 2014). It is then possible to detect nanometric impurities that can significantly influence the dielectric properties of the sample under investigation.

The area under test has a rich Raman spectrum with the peaks at 225, 292, 410 and 612 cm$^{-1}$ previously reported in the literature (https://www.fis.unipr.it/phevix/ramandb.php?plot=Hematite1&submit=Go).

Subsequently, a Raman spectrum obtained by sum-averaging spectra at 7 positions on an inclusion of the iron oxide, was compared to the spectra obtained by analysing a sample of pure hematite mineral and published Raman spectra (https://www.fis.unipr.it/phevix/ramandb.php?plot=Hematite1&submit=Go). Results are reported in Figure 8.

The Raman spectrum of $\alpha$-Fe$_2$O$_3$, hematite, exhibits seven peaks which were previously assigned to A$_{1g}$ mode (225 and 498 cm$^{-1}$) and five peaks of E$_g$ modes (247, 293, 299, 412 and 613 cm$^{-1}$ peaks) (Porto and Krishnan, 1967; Beattie and Gibson, 1970).

In both the inclusion and the pure hematite sample we observed a broad band at about 1300 cm$^{-1}$ (not shown on Figure 8) which was previously assigned to the collective spin movement (two-magnon scattering) which is very sensitive to temperature (Thibeau, 1978). We kept the condition of the Raman acquisition such to minimize any heating effect, which is known to shift peak positions as well as cause peak broadening.

Slight differences in the Raman spectrum of the inclusion compared to the pure hematite spectrum, especially in the relative intensity of the two major A1g and Eg bands at 225 and 293 cm$^{-1}$ stems from the unknown orientation of the inclusion crystal lattice in respect to the laser polarization. Also it allows the identification of the inclusion as high quality hematite.

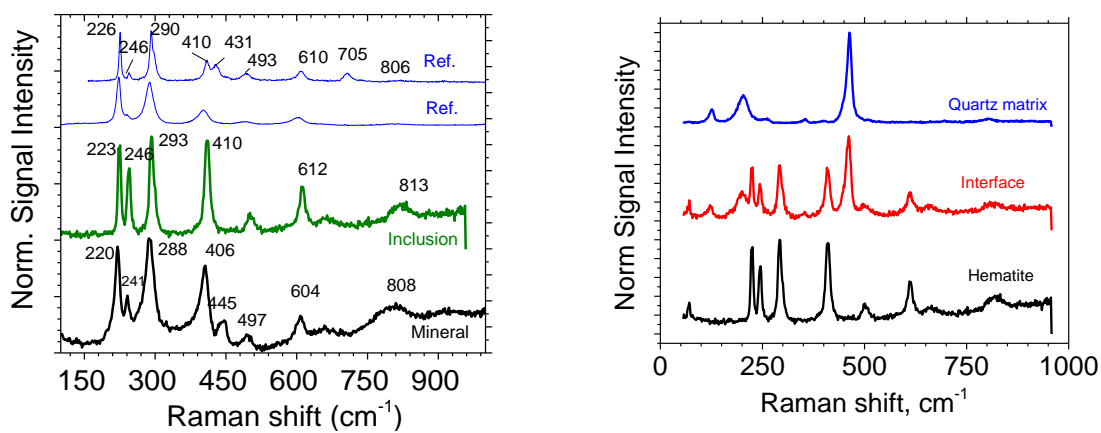

*Fig.8 a) Raman spectra of hematite inclusion (green), powder of pure minerals analysed by the same system (black), which are the sum-average of seven spectra, compared with the published Raman spectra of hematite, b) Raman spectra of quartz, quartz- hematite interface and hematite suggesting that the interface is a mixture of two pure minerals (quartz and hematite). The quartz Raman spectrum suggests that SiO2 is present in the $\alpha$–quartz polymorph based on previous assignment by Kingma and Hemley (1994).*

*4.2. Scanning Microwave Microscopy Analysis*

The SMM technique, described in the previous section, has been applied to the hematite minerals, after their preliminary chemical and elemental characterization.

In order to quantify the dielectric constant of the ore a calibration procedure (as described in Section 2.4.) was applied before and after the measurements. Five materials with known dielectric constant were measured and the results were interpolated in the calibration curve in Figure 5.

The microscopic image of the dielectric constant, namely sMIM-C described in Section 2.3, of the hematite phase is shown in Figure 9. The average dielectric contrast between the microwave lossy inclusion and the surrounding matrix is also imaged. A further difference can be appreciated between the hematite (yellow), the calcite (light purple) and the quartz (dark purple). The location and shape of the different elements from the SEM-BSE scan (Fig.7) are clearly visible in the dielectric contrast image shown in Figure 9.

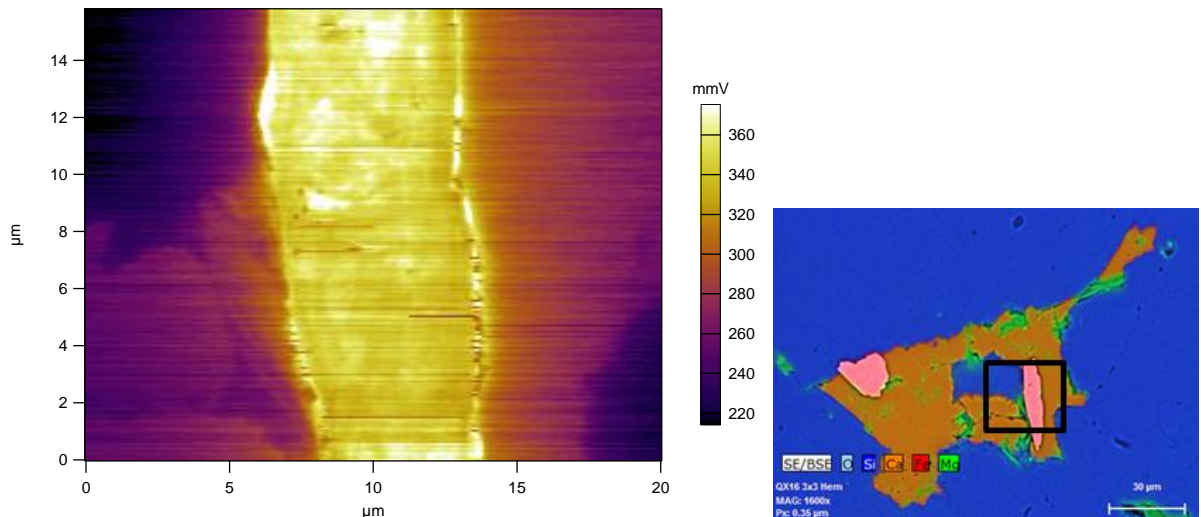

*Fig. 9 Map of the ΔC data (sMIM-C) of the area under analysis [left]. The dielectric contrast between hematite (dark purple), calcite (light purple) and quartz (yellow) is clearly visible. This picture highlights the dielectric contrast between the materials. On the right panel, the black square indicates the area of the sample analysed by the SMM.*

It is then possible to convert the raw data coming from the C-channel of the microscope to real values of the dielectric constant and then to have a nanometric scale, quantitative map of the values (Fig.10) through the calibration measurements performed before and after each experiment. Quartz, calcite and hematite respectively are identified by different colour tones (blue, light blue, green/yellow) that correspond to different values of $\varepsilon'$.

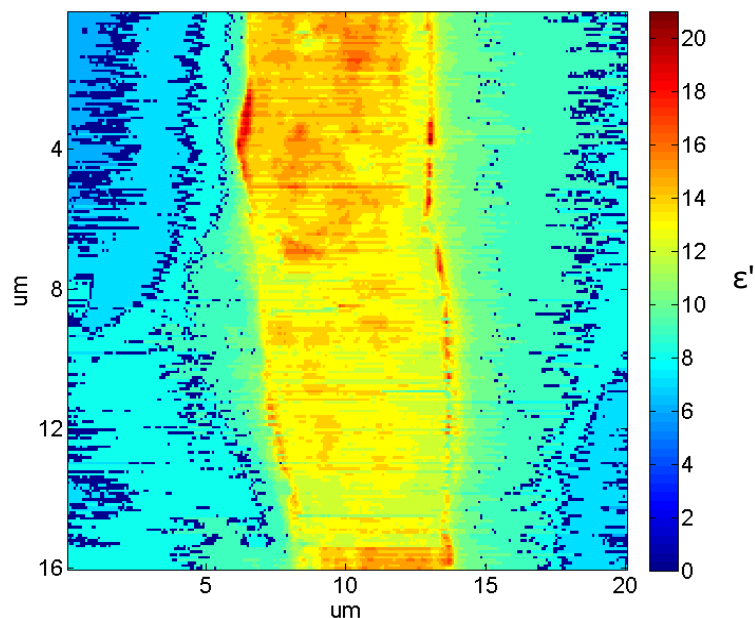

*Fig.10 Map of the dielectric constant $\varepsilon'$ of the area under analysis. Quartz in the top-left corner and bottom-right one is blue ($\varepsilon' = [4-5]$), calcite around the elongated phase is light blue ($\varepsilon' = [8-9]$),*

*central hematite is green-yellow ($\varepsilon' = [12 - 15]$) The elongated artefacts are due to the occasional imperfect contact between tip and sample during the scanning. They are also visible in Figure 9.*

Finally, the loss tangent and conductivity of the hematite were evaluated. The proportionality constant $\kappa$ described above was obtained by the numerical model implemented in COMSOL Multiphysics© and discussed in Section 3. From the literature we estimated the conductivity of the hematite (Nelson et al., 1989; Peng et al., 2012; Hotta et al., 2010). to be within the range $\sigma = [0.1 - 1] S/m$. From the dielectric constant experimental evaluation the range $\varepsilon' = [12 - 15]$ was assumed for the hematite inclusion; from these assumptions $\kappa = [8 - 11]$ was derived.

For the loss tangent of the hematite inclusion, shown in Fig.10, values of $\varepsilon'' = [4 - 9.4]$ and $\sigma = [0.8 - 2.2] S/m$ were obtained. Results for the imaginary part of the complex dielectric permittivity ($\varepsilon''$), for $tan\delta$ as from (3), and for the conductivity σ of the inclusion are reported in Figure 11. The surrounding gangue material is microwave lossless with $\varepsilon'' = 0$ and $\sigma = 0$.

These data are in good agreement with earlier literature results, taking into account a wide variability of conditions and differences between the measurement methods. For example, in Nelson et al. (1989), the dielectric values for the bulk density of the material were derived from theoretical calculation, while the measurements were performed on powders of pure minerals at different densities. In Peng et al. (2012) the complex dielectric permittivity data were measured by cavity perturbation technique using hematite samples composed of powders of pure minerals with a bulk density approaching to the solid. In Hotta et al. (2010) the open-ended coaxial probe technique was used on compressed powders of pure hematite crystals with different densities.

Even if the topography of the hematite phase is absolutely flat, as shown in the AFM image of Figure 12, a certain variability along its surface is reported. Several spots in the hematite show a higher conductivity (red spots) than the rest of the phase. However, such variability is not reported as chemical impurities in the SEM/EDX. In addition, the purity of the mineral is confirmed by the Raman analysis.

The variability in the conductivity can be ascribed to either impurities with low atomic weight at low concentration or as an intrinsic property of the hematite itself.

In both cases, as the increase in conductivity is highly localized, it cannot be taken into account in an average 'bulk' dielectric measurement, as in Nelson et al. (1989), Peng et al. (2012) and Hotta et al. (2010).

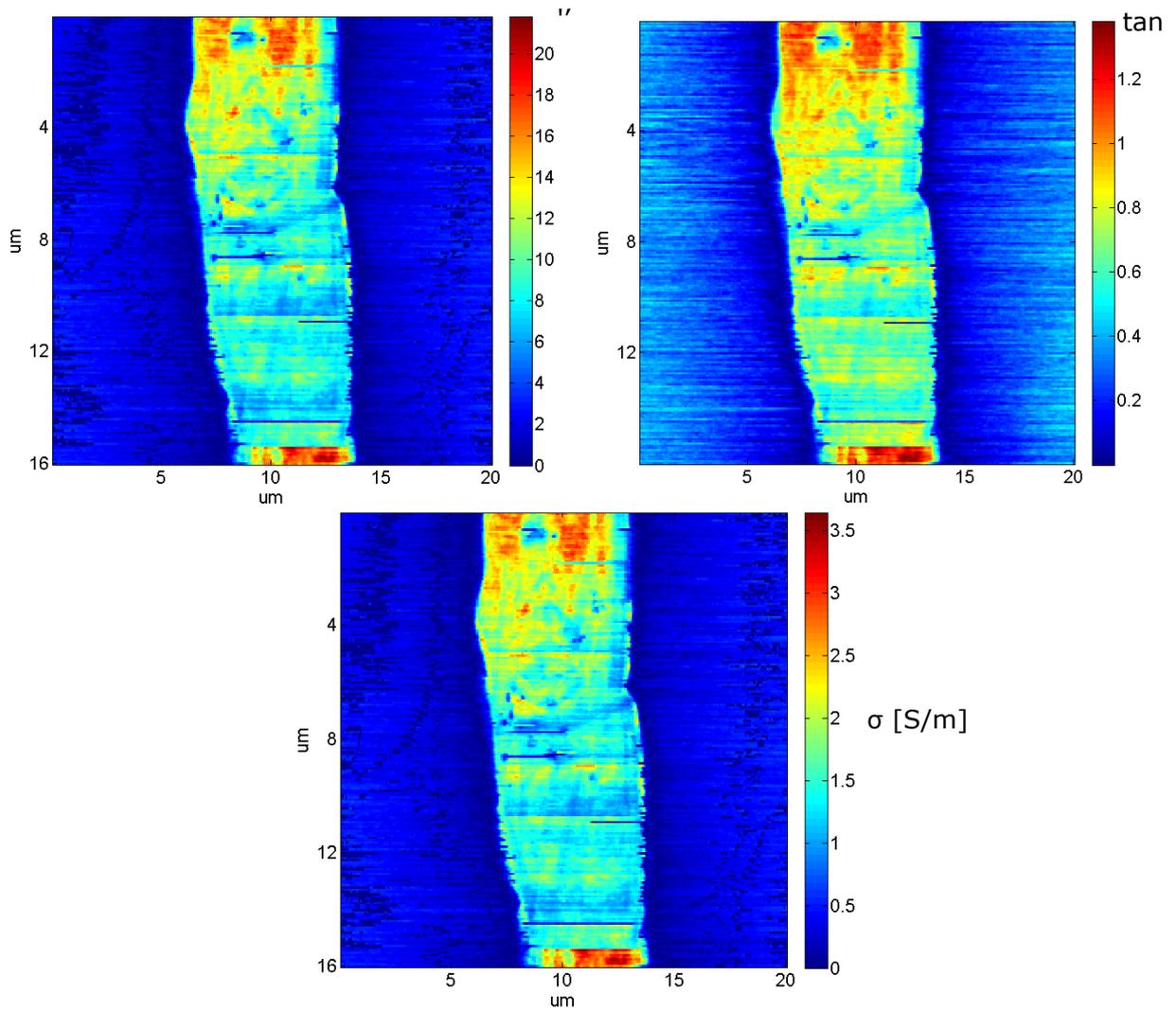

Fig.11 Maps of the ε", tanδ and σ of the area under test. Note a few spots on top and bottom of the inclusion exhibit a slightly higher conductivity, even if the hematite is chemically pure (Fig.7) and the topographically is flat (Fig.12).

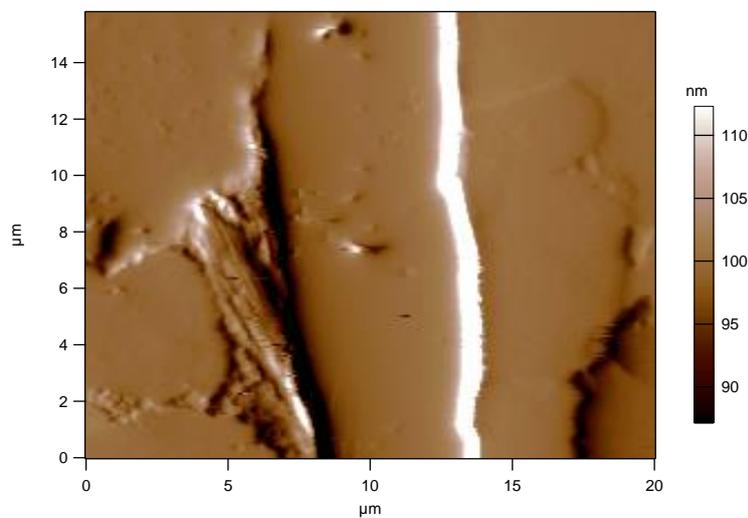

Fig.12 AFM topography of the same area under test. The scanned surface is flat within few nm. The hematite surface is flat with few pinch defects on its surface not clearly recognizable in the previous dielectric maps.

The reproducibility of sMIM measurements was demonstrated by testing the same inclusions with different tips and associated sets of calibration curves. Furthermore, the second hematite inclusion detected by the SEM/EDX analysis has been tested using sMIM with similar results ($\varepsilon' = 12$ and $\varepsilon'' = 5$ with $\sigma = 0.83\ S/m$).

Conclusions

In this work we present a nanoscale dielectric characterization of a mineral inclusion in a complex natural rock. In particular, a micrometric-sized inclusion of hematite, embedded in a gangue matrix, has been analysed by a scanning microwave microscope and details about its dielectric constant, losses and conductivity are presented.

All data are in good agreement with previous literature for pure mineral samples. The dielectric constant of the hematite inclusion was in the range 12 to 15 while the conductivity was estimated to be between 0.8 and 2.2 S/m. Dielectric constants of the calcite and quartz have been mapped as well; their values are in the range 8 to 9 and 4 to 5 respectively.

The results from the scanning electron microscopy/energy-dispersive x-ray spectroscopy analysis and micro Raman analysis are in agreement with the dielectric map obtained by the scanning microwave microscopy analysis. The hematite grain is chemically pure and the related dielectric map is essentially uniform. However, a certain variability inside the grain itself only a few tens of micrometres in size was noted and spatially mapped.

The capability of the scanning microwave microscope for measurement of the dielectric properties of a mineral inclusion in a real rock, as opposed to 'bulk' dielectric measurement techniques has been demonstrated.

The variations recorded in the conductivity can be ascribed to either impurities with low molecular weight at low concentration or to intrinsic variability in the hematite itself.

In order to confirm these hypotheses, more accurate chemical characterization techniques have to be applied (*e.g.* electron microprobe).

In this way the scanning microwave microscope can be efficiently used for quantification of the dielectric properties of inclusions within mineral samples and also to relate such properties to their physical structure, phases and chemical composition at a nanometric scale.

Acknowledgement

Microwave microscopy and confocal micro-Raman spectroscopy measurements were performed at Center for Nanophase Materials Sciences, which is a Department of Energy Office of Science User Facility.